\documentclass[aps,twocolumn,groupedaddress,amsmath,amssymb]{revtex4}
\usepackage{amsmath, dsfont}
\usepackage{amssymb}
\usepackage[dvips]{graphics}
\usepackage{epsfig}
\usepackage{calc}
\usepackage{amsfonts}
\usepackage{graphicx}
\usepackage[ansinew]{inputenc}

\begin{document}
\setcounter{page}{1}
\title{Towards the Cardy formula for hyperscaling violation black holes}

\author{Mois\'es~Bravo-Gaete, Sebasti\'an~G\'omez and Mokhtar~Hassa\"ine}
\email{mbravog-at-inst-mat.utalca.cl, sebago-at-inst-mat.utalca.cl,
hassaine-at-inst-mat.utalca.cl} \affiliation{Instituto de
Matem\'atica y F\'isica, Universidad de Talca, Casilla 747, Talca,
Chile}

\begin{abstract}
The aim of this paper is to propose a generalized Cardy formula in
the case of three-dimensional hyperscaling violation black holes. We
first note that for the hyperscaling violation metrics, the scaling
of the entropy in term of the temperature (defined as the effective
spatial dimensionality divided by the dynamical exponent) depends
explicitly on the gravity theory. Starting from this observation, we
first explore the case of quadratic curvature gravity theory for
which we derive four classes of asymptotically hyperscaling
violation black holes. For each solution, we compute their masses as
well as those of their soliton counterparts obtained through a
double Wick rotation. Assuming that the partition function has a
certain invariance involving the effective spatial dimensionality, a
generalized Cardy formula is derived. This latter is shown to
correctly reproduce the entropy where the ground state is identified
with the soliton. Comparing our formula with the one derived in the
standard Einstein gravity case with source, we stress the role
played by the effective spatial dimensionality. From this
observation, we speculate the general form of the Cardy formula  in
the case of hyperscaling violation metric for an arbitrary value of
the effective spatial dimensionality. Finally, we test the viability
of this formula in the case of cubic gravity theory.
\end{abstract}

\maketitle

%%%%%%%%%%%%%%%%%%%%%%%
\section{Introduction}
%%%%%%%%%%%%%%%%%%%%%%%
In the last decade, there are been  a real interest in extending the
ideas underlying the AdS/CFT correspondence \cite{Maldacena:1997re}
in order to gain a better understanding of strongly condensed matter
systems. In condensed matter physics, a quantum phase transition is
a transition that occurs between two different phases at zero
temperature. At this critical point, the system may enjoy an
anisotropic scaling symmetry or even display an hyperscaling
violation which is reflected by the fact that the entropy $S$ scales
with respect to the temperature $T$ as
\cite{Gouteraux:2011ce,Sachdev:2012dq,Charmousis:2012dw}
\begin{eqnarray}
S\sim T^{\frac{d_{\tiny{\mbox{eff}}}}{z}}, \label{scaleST}
\end{eqnarray}
where $d_{\tiny{\mbox{eff}}}$ is the effective spatial
dimensionality, and where $z$ is the dynamical exponent responsible
of the anisotropy of the system
\begin{equation}\label{eq:scaling}
t\mapsto\lambda^z\,t,\qquad \vec{x}\mapsto\lambda\vec{x}.
\end{equation}
In the non-relativistic version of the AdS/CFT correspondence, the
gravity dual metric in the anisotropic case is played by the
Lifshitz metric whose line element in $D$ dimensions is given by
\cite{Kachru:2008yh}
\begin{equation}\label{eq:Lifshitzmetric}
ds_{\mathrm{L}}^2=-\frac{r^{2z}}{l^{2z}}dt^2+\frac{l^2}{r^2}dr^2
+\frac{r^2}{l^2}d\vec{x}^2,
\end{equation}
where $\vec{x}=(x^1,\ldots,x^{D-2})$. For the Lifshitz metric, it is
easy to see that the anisotropic transformations (\ref{eq:scaling})
together with $r\to \lambda^{-1} r$ act as an isometry. It is
well-known also that, in contrast to the AdS isotropic case $z=1$,
Lifshitz metrics or their black hole extensions are not solutions of
the standard Einstein gravity but instead require the introduction
of some extra matter source and/or to consider higher-order gravity
theories, see e. g.
\cite{Taylor:2008tg,AyonBeato:2009nh,Pang:2009pd,Maeda:2011jj,Ayon-Beato:2015jga,Bravo-Gaete:2015xea,Correa:2014ika,Alvarez:2014pra}.
Moreover, in the Lifshitz case, the effective spatial dimensionality
has a fixed value which depends on the dimension as
$d_{\tiny{\mbox{eff}}}=D-2$.

On the other hand, systems which display an hyperscaling are
described by the so-called hyperscaling violation metric whose line
elements is conformally related to the Lifshitz metric as
\begin{eqnarray}
ds^{2}&=&\frac{1}{r^{\frac{{2\theta}}{D-2}}}\Big[-r^{2z}
{dt^2}+\frac{dr^2}{r^2 }+r^2d\vec{x}^2\Big], \label{HSV}
\end{eqnarray}
where now the transformations (\ref{eq:scaling}) together with $r\to
\lambda^{-1} r$ act rather like a conformal transformation, $ds^2\to
\lambda^{2\theta/(D-2)}ds^2$. Examples of hyperscaling violation
black holes are known in the literature, see e. g.
\cite{Dong:2012se,Alishahiha:2012qu,HVMBH1,HVMBH2,Shaghoulian:2015dwa,Hassaine:2015ifa}.
As shown below, in the hyperscaling violation case, the effective
spatial dimensionality will explicitly involve the hyperscaling
violation exponent $\theta$, and this dependence will vary in
function of the gravity theory considered. For example, in the case
of standard three-dimensional Einstein gravity with source, we have
$d_{\tiny{\mbox{eff}}}=1-\theta$, see e. g.
\cite{Shaghoulian:2015dwa} while (see below) for quadratic (resp.
cubic) gravity theories in three dimensions we will have
$d_{\tiny{\mbox{eff}}}=1+\theta$ (resp.
$d_{\tiny{\mbox{eff}}}=1+3\theta$).

The aim of this paper is to speculate the form of a generalized
Cardy formula in three dimensions in the case of hyperscaling
violation metric under the assumption that the ground state is
identified with the soliton. This task has been realized  in the
Lifshitz case by exploiting an isomorphism between the Lifshitz Lie
algebras with dynamical exponent $z$ and $z^{-1}$ in two dimensions,
see \cite{Gonzalez:2011nz}. In this last reference, the Cardy
formula has been derived assuming that the ground state is
identified with the soliton which is, in addition, separated from
the black hole spectrum by a gap. Note that in Ref.
\cite{Ayon-Beato:2015jga}, we have tested the viability of the Cardy
formula of \cite{Gonzalez:2011nz} in some concrete examples of
Lifshitz black holes with a source given by a scalar field
nonminimally coupled. Just to be complete, it is worth mentioning
that in the standard three-dimensional AdS gravity supported by
scalar fields, the identification of the solitons as ground states
turns out to be essential for microscopically counting for the black
holes entropy using Cardy formula \cite{Correa:2010hf}.

From now, one can anticipate that, in contrast with the
three-dimensional Lifshitz case where the effective spatial
dimensionality always takes the value  $d_{\tiny{\mbox{eff}}}=1$,
our proposal for the Cardy formula in the case of hyperscaling
violation metric will instead depend explicitly on the exponent
$\theta$, and this dependence is inheriting from the theory
considered.  In order to guess the form of the possible Cardy
formula for hyperscaling violation, we will consider the case of
quadratic and cubic curvature gravity theories for which we will
derive black hole and their  soliton counterpart solutions. The
black hole and soliton masses will be computed using the quasilocal
formalism introduced in \cite{Kim:2013zha,Gim:2014nba} where in the
next section we will briefly recall the main lines. Comparing our
results obtained in the quadratic and cubic cases to those derived
recently in \cite{Shaghoulian:2015dwa} in the case of standard
Einstein gravity with source, we will speculate the form of the
generalized Cardy formula with  a ground state identified with the
soliton. This generalized formula which reduces to the one derived
in \cite{Shaghoulian:2015dwa} in the standard Einstein case will be
shown to depend explicitly on the effective spatial dimensionality,
and we will show that this formula correctly reproduces the
semiclassical entropy in the different examples exploited in this
paper.

The plan of the paper is organized as follows. In the next section,
we consider the most general quadratic curvature in three dimensions
for which we derive four classes of hyperscaling violation black
holes. For each solution, we compute their masses as well as well
those of their respective soliton counterparts. We also show that
there always exists an election of the coupling constant that
ensures the black hole masses to be positive while those of their
soliton counterparts turn to be negative; this ensures the existence
of  a gap in the spectrum of the solutions. In addition, assuming
that the theory is invariant under a certain modular transformation
whose form depends on the effective spatial dimensionality, we will
be able to obtain a Cardy formula. This latter under the assumption
that the ground state is identified with the soliton, correctly
reproduces the expressions of the Wald entropy for each of our
examples. Then, comparing our formula with the one derived recently
in the standard Einstein case with source
\cite{Shaghoulian:2015dwa}, we propose a generalized Cardy formula
whose expression depends explicitly on the effective spatial
dimensionality. In the last section, we check the validity of this
Cardy formula in the case of the most general cubic gravity theory
in three dimensions. Finally, the last section is devoted to our
conclusions.

%%%%%%%%%%%%%%%%%%%%%%%%%%%%%%%%%%%%%%%%%%%%%
\section{Quadratic curvature gravity theory}
%%%%%%%%%%%%%%%%%%%%%%%%%%%%%%%%%%%%%%%%%%%%%
In three dimensions, we consider the most general quadratic
curvature gravity theory given by the following action
\begin{eqnarray}\label{action}
{S}[g_{\mu \nu}]&=&\int d^{3}x\sqrt{-g}\left( \beta_1 R^{2}+\beta_2
R_{\mu \nu }R^{\mu \nu}\right) \nonumber \\
&=&\int d^{3}x \sqrt{-g} {\cal L},
\end{eqnarray}
where $\beta_1$ and $\beta_2$ are two coupling constants. The field
equations arising from the variation of the action (\ref{action})
read
\begin{eqnarray}
\beta_2\square{R}_{\mu\nu}
+\frac12\left(4\beta_1+\beta_2\right)g_{\mu\nu}\square{R}
-\left(2\beta_1+\beta_2\right)\nabla_\mu\nabla_\nu{R}
\nonumber\\
+2\beta_2R_{\mu\alpha\nu\beta}R^{\alpha\beta}
+2\beta_1RR_{\mu\nu}\nonumber\\
-\frac12\left(\beta_1{R}^2+\beta_2{R}_{\alpha\beta}{R}^{\alpha\beta}
\right)g_{\mu\nu}=0.\label{eq:squareGrav}
\end{eqnarray}
Since we are looking for asymptotically hyperscaling violation black
holes, we posit the following ansatz for the metric
\begin{eqnarray}
ds^{2}&=&\frac{1}{r^{{2\theta}}} \Big[-r^{2z} f(r)
{dt^2}+\frac{dr^2}{r^2 f(r)}+r^2 d\varphi^{2}\Big],
\label{HSVmetricf}
\end{eqnarray}
where the metric function $f(r)$ possesses at least one root, and
satisfies $\lim_{r\to\infty}f(r)=1$ in order to reproduce
asymptotically the hyperscaling violation metric (\ref{HSV}). To
simplify the computations, we will consider the following particular
form for the metric function
\begin{eqnarray}
f(r)=1-\left(\frac{r_h}{r}\right)^{\alpha}, \label{metricf}
\end{eqnarray}
where $\alpha$ is a constant with $\alpha>0$, and where $r_h$
denotes the location of the horizon. In fact, there exist more
general solutions within the ansatz (\ref{HSVmetricf}) which involve
more than an one integration constant. We will discuss one of these
solutions at the end of the section devoted to the discussion.

In addition, since we will be interested on the thermodynamic
properties of the solutions, we compute the Wald entropy for the
generic solution (\ref{HSVmetricf}-\ref{metricf}) which is given by
\begin{eqnarray}
{\cal S}_{W}&=&8\,\pi^2 \,\alpha\,(r_h)^{1+\theta}\Big[\left(
8\,\theta- 6\,z+2\,\alpha-4 \right) {\beta_1}\nonumber\\
&+& \left( 3\,\theta-3\,z-1+\alpha \right) {\beta_2} \Big],
\label{waldentrop}
\end{eqnarray}
while the Hawking temperature takes the form
\begin{equation}
T={\frac {\alpha\,(r_h)^{z}}{4\,\pi }}. \label{temp}
\end{equation}
It is important to stress  from now that the entropy scales with the
temperature as
\begin{eqnarray}
{\cal S}_{W}\sim T^{\frac{1+\theta}{z}}, \label{scaleq2}
\end{eqnarray}
which means that the effective spatial dimensionality
$d_{\tiny{\mbox{eff}}}$ as defined in (\ref{scaleST}) is given by
\begin{eqnarray}
d_{\tiny{\mbox{eff}}}=1+\theta. \label{deff}
\end{eqnarray}
This is due to the fact that we are considering quadratic curvature
gravity theories. Indeed, in the case of standard Einstein-Hilbert
gravity supplemented by  a source given by the (Maxwell)-dilaton
Lagrangian \cite{Dong:2012se,Alishahiha:2012qu,Shaghoulian:2015dwa},
the effective spatial dimensionality is instead
$d_{\tiny{\mbox{eff}}}=1-\theta$. From now, one can already conclude
to the non-universal character of the scaling of the entropy in term
of the temperature in the case of hyperscaling violation metric. In
other words, this means that the effective spatial dimensionality
will depend on the theory considered. This is in contrast with the
Lifshitz case $\theta=0$ where, independently of the theory, the
entropy always scales as ${\cal S}_{W}\sim T^{\frac{1}{z}}$, see \cite{Gonzalez:2011nz} in the general context and 
\cite{AyonBeato:2009nh}, 
\cite{Ayon-Beato:2015jga,Bravo-Gaete:2015xea,Correa:2014ika} for concrete examples. This
remark will have its importance when speculating the form of the
generalized Cardy in the case of hyperscaling violation metric.

In order to compute the mass of our black hole and soliton
solutions, we opt for the quasilocal formalism presented in
\cite{Kim:2013zha,Gim:2014nba} whose main result lies in the
relation established between the the off-shell ADT potential
$Q_{\mbox{\tiny{ADT}}}^{\mu\nu}$  and the off-shell Noether
potential $K^{\mu\nu}$ in the form
\begin{eqnarray}
\sqrt{-g}\,Q_{\mbox{\tiny{ADT}}}^{\mu\nu}=\frac{1}{2}\delta
K^{\mu\nu}-\xi^{[\mu}\Theta^{\nu]}, \label{reladtpot}
\end{eqnarray}
where $\xi^{\mu}\partial_{\mu}$ denotes the Killing vector field
which in our case is $\partial_t$, and $\Theta^{\mu}$ represents a
surface term arising from the variation of the action. More
precisely, the expressions appearing in (\ref{reladtpot}) are given
by
\begin{eqnarray}
\Theta^\mu &=&2\sqrt{-g}\Big[P^{\mu(\alpha
\beta)\gamma}\nabla_\gamma\delta g_{\alpha\beta} -\delta
g_{\alpha\beta}\nabla_\gamma
P^{\mu(\alpha\beta)\gamma}\Big], \label{eq:theta}\\
K^{\mu\nu}
&=&\sqrt{-g}\,\Big[2P^{\mu\nu\rho\sigma}\nabla_\rho\xi_\sigma
-4\xi_\sigma\nabla_\rho P^{\mu\nu\rho\sigma}\Big], \label{eq:K}
\end{eqnarray}
with $P^{\mu \nu \sigma \rho}=\frac{\partial {\cal L}}{\partial
R_{\mu \nu \sigma \rho}}$, where ${\cal L}$ is the Lagrangian
defined in (\ref{action}). Since the ansatz metric
(\ref{HSVmetricf}-\ref{metricf}) depends continuously on the
integration constant ${r_h}^{\alpha}$, one can define the conserved
charge associated to the Killing field $\partial_t$ (which
corresponds to the mass) in the interior region and not in the
asymptotic region by introducing a parameter $s$ with range $s\in
[0,1]$ as $s \, {r_h}^{\alpha}$. The advantage of this re-definition
lies in the fact that it allows to interpolate between the free
parameter solution $s=0$ and the solution with $s=1$. In doing so,
the quasi-local charge is defined as \cite{Kim:2013zha,Gim:2014nba}
\begin{equation}
\label{eq:charge} {\cal M}_{\tiny{\mbox{bh}}}(\xi)\!=\int_{\cal B}\!
dx_{\mu\nu}\Big(\Delta K^{\mu\nu}(\xi)-2\xi^{[\mu} \!\! \int^1_0ds~
\Theta^{\nu]}(\xi\, |\, s)\Big),
\end{equation}
where $\Delta K^{\mu\nu}(\xi) \equiv
K^{\mu\nu}_{s=1}(\xi)-K^{\mu\nu}_{s=0}(\xi)$ denotes the variation
of the Noether potential from the vacuum solution, and $dx_{\mu\nu}$
represents the integration over the compact co-dimension
two-subspace. For the generic solution
(\ref{HSVmetricf}-\ref{metricf}), after a tedious but
straightforward computation, this last expression becomes
parameterized as
\begin{eqnarray}
\label{eq:mastereq} {\cal M}_{\tiny{\mbox{bh}}}=2\pi\Big[
(r_h)^{\alpha} \Psi_{1}{r}^{1+\theta+z-\alpha}+ (r_h) ^{2\,\alpha}
\Psi_{2}{r}^{1+\theta+z-2\,\alpha} \Big],&&
\end{eqnarray}
where $\Psi_{1}$ and $\Psi_{2}$ are constants given by
\begin{eqnarray*}
\Psi_{1}&=&\Big[ 4\,{\alpha}^{3}+ \big( -8+4\,\theta-8\,z
\big) {\alpha}^{2}+ \big( -40\,{\theta}^{2}+8\,\theta\nonumber\\
&+&8 -4\,{z}^{2} +36\,z\theta \big) \alpha+4\, \big(
2\,z-1-7\,\theta
\big)\big( {z}^{2}-2\,z\theta\nonumber\\
&+&z+1+{\theta}^{2}-2\,\theta \big) \Big] {\beta_1}+\Big[
2\,{\alpha}^{3}+ \big( -4\,z-3+\theta \big) {\alpha}^{2}\nonumber\\
&+& \big( -15\,{\theta}^{2}+15\,z\theta+3\,z+2\,\theta-2\,{z}^{2}+1
\big) \alpha-2\,{z}^{2}\nonumber\\
&-&18\,{z}^{2}\theta-6\,\theta -2-12\,z\theta+4
\,{z}^{3}+24\,z{\theta}^{2}-10\,{\theta}^{3}\nonumber\\
&+&18\,{\theta}^{2} +4\,z \Big] {\beta_2},
\nonumber\\ \nonumber\\
\Psi_{2}&=& \Big[ -2\,{\alpha}^{ 3}+ \big( 4\,z+5-\theta \big)
{\alpha}^{2}+ \big( 24\,{\theta}^{2} -2\,\theta-21\,z\theta
\nonumber\\
&-&6+2\,{z}^{2}-3\,z \big) \alpha-2\, \big( 2\, z-1-7\,\theta \big)
\big( {z}^{2}-2\,z\theta+z\nonumber\\
&+&1+{\theta}^{2}-2\, \theta \big) \Big] {\beta_1}+\Big[
-{\alpha}^{3}+ \big( 2+2\,z \big) {\alpha}^{2}+ \big( -9\,
z\theta\nonumber\\
&+&{z}^{2}-3\,z-1+9\,{\theta}^{2} \big) \alpha+9\,{z}^{2}\theta
+1+3\,\theta+6\,z\theta+{z}^{2}\nonumber\\
&-&2\,z-12\,z{\theta}^{2}-2\,{z}^{3}-9\,{
\theta}^{2}+5\,{\theta}^{3} \Big] {\beta_2}.
\end{eqnarray*}
Since this expression of the mass must be independent of the radial
coordinate $r$, the non-zero black hole solutions, as shown below,
will automatically satisfy $\alpha=1+\theta+z$ with $\Psi_{2}=0$ or
$\alpha=(1+\theta+z)/2$ with $\Psi_{1}=0$.

Our aim being to derive a generalized version of the Cardy formula
for the hyperscaling violation metric where the ground state is
played by the soliton, we note that for a black hole solution of the
form (\ref{HSVmetricf}-\ref{metricf}), its soliton counterpart
obtained through a double Wick rotation will have the 
following generic form
\begin{eqnarray}
ds^{2}&=&\frac{1}{r^{{2\theta}}} \Big[-r^{2} {dt^2}+\frac{dr^2}{r^2
\tilde{f}(r)}+r^{2z}\tilde{f}(r)
d\varphi^{2}\Big], \label{HSVmetricsoliton}\\
\tilde{f}(r)&=&1-\left(\frac{\tilde{r_h}}{r}\right)^{\alpha} ,
\nonumber
\end{eqnarray}
where we have defined
$$
\tilde{r_h}=\left(\frac{2}{\alpha}\right)^{\frac{1}{z}}.
$$
Along the same lines as before, the quasilocal mass for the soliton
(\ref{HSVmetricsoliton}) can schematically be written as
\begin{eqnarray}
{\cal M}_{\tiny{\mbox{sol}}}= 2\pi\Big[ (\tilde{r_h})^{\alpha}
\Phi_{1}{r}^{1+\theta+z-\alpha}+ (\tilde{r_h}) ^{2\,\alpha}
\Phi_{2}{r}^{1+\theta+z-2\,\alpha} \Big],&& \label{solitonicmass}
\end{eqnarray}
where $\Phi_{1}$ and $\Phi_{2}$ read
\begin{eqnarray*}
\Phi_{1}&=&\Big[4\,{\alpha}^{3}+ \big( -12\,z-4+4\,\theta \big)
{\alpha}^{2}+\big( 4+12\,{z}^{2}+12\,\theta\,z\nonumber\\
&+&16\,\theta-36\,{\theta}^{2} \big) \alpha-4\, \big(
2\,z-3+7\,\theta \big)  \big( {z}^{2}-2 \,\theta\,z+z\nonumber\\
&+&1 -2\,\theta+{\theta}^{2} \big) \Big]\,\beta_1+\Big[{\alpha}^{3}+
\big( 2\,\theta-3\,z-2 \big) {\alpha}^{2}+ \big( 6\,
\theta\nonumber\\
&+&3\,\theta\,z +4\,{z}^{2}-13\,{\theta}^{2}+3-z \big) \alpha+6-4
\,{z}^{3}-12\,\theta\,z\nonumber\\
&-&22\,\theta+6\,{z}^{2}-10\,{\theta}^{3}-4\,z-2
\,{z}^{2}\theta+16\,{\theta}^{2}z\nonumber\\
&+&26\,{\theta}^{2} \Big]\,\beta_2,
\nonumber\\
\nonumber\\
\Phi_{2}&=& \Big[-4\,{\alpha}^{3}+ \big( 14\,z+5-9\,\theta \big)
{\alpha}^{2}+\big( -14\,{z}^{2}-10\,\theta\nonumber\\
&-&2+20\,{\theta}^{2}+3\,\theta\,z-3\,z
\big) \alpha+2\, \big( 2\,z-3+7\,\theta \big)  \big( {z}^{2}\nonumber\\
&-&2\,\theta\,z+z+1-2\,\theta+{\theta}^{2} \big)
\Big]\,\beta_1+\Big[-{\alpha}^{3}+ \big(
-4\,\theta\nonumber\\
&+&4\,z+2 \big) {\alpha}^{2}+ \big( -
3-5\,{z}^{2}+z+3\,\theta\,z+7\,{\theta}^{2}-4\,\theta \big)
\alpha\nonumber\\
&-&3 +2\,{z}^{3}+6\,\theta\,z
+11\,\theta-3\,{z}^{2}+5\,{\theta}^{3}+2\,z+{z
}^{2}\theta\nonumber\\
&-&8\,{\theta}^{2}z-13\,{\theta}^{2} \Big]\,\beta_2.
\end{eqnarray*}

%%%%%%%%%%%%%%%%%%%%%%%%%%%%%%%%%%%%%%%%%%%%%%%%%%%%%%%%%%%%%%%%%%%%%%%%%%%%%%%%%
\subsection{Four classes of black hole solutions and their soliton counterparts}
%%%%%%%%%%%%%%%%%%%%%%%%%%%%%%%%%%%%%%%%%%%%%%%%%%%%%%%%%%%%%%%%%%%%%%%%%%%%%%%%%
In what follows, we present four classes of black hole solutions of
the field equations (\ref{eq:squareGrav}) within the ansatz given by
(\ref{HSVmetricf}-\ref{metricf}). For each solution, we compute the
mass (\ref{eq:charge}) as well as the mass of their soliton
counterpart (\ref{HSVmetricsoliton}) through the generic expression
given by (\ref{solitonicmass}). We check that the first law of black
hole thermodynamics
\begin{equation}\label{firstlaw}
d {\cal M}_{\tiny{\mbox{bh}}}=T d {\cal S}_{W},
\end{equation}
is valid in these four cases.

The first family of solutions is obtained for an hyperscaling
violation exponent and dynamical exponent that take the values
$\theta=2$ and $z=1$, and the line element is given by
\begin{eqnarray}
ds^{2}&=&\frac{1}{r^{4}}\,\Big[-r^{2} f(r) {dt^2}+\frac{dr^2}{r^2
f(r)}+
r^{2} d\varphi^{2}\Big], \label{HSVmetricfsoln4}\\
f(r)&=&1-\left(\frac{r_h}{r}\right)^{4} , \nonumber
\end{eqnarray}
provided that the coupling constants $\beta_1$ and $\beta_2$ are
tied as
\begin{eqnarray}\label{beta1soln3}
\beta_1=-\frac{5}{13}\,\beta_2.
\end{eqnarray}
In this case, the Wald entropy ${\cal S}_{W}$(\ref{waldentrop}), the
Hawking temperature $T$ (\ref{temp}) and the black hole mass ${\cal
M}_{\tiny{\mbox{bh}}}$ (\ref{eq:mastereq}) are given by
\begin{eqnarray}
{\cal S}_{W}&=&\frac{256}{13}\,\pi^{2}\,\beta_2 \,(r_h)^{3},\qquad
T=\frac{r_h}{\pi},\nonumber\\
{\cal M}_{\tiny{\mbox{bh}}}&=&\frac{192}{13}\,\pi\,\beta_2\,(r_h)^4,
\label{entemmasssol4}
\end{eqnarray}
and it is a matter of check to see that the first law
(\ref{firstlaw}) is satisfied. The corresponding soliton whose line
element (\ref{HSVmetricsoliton}) written in terms of the "regular"
coordinates
$$
r=\frac{1}{2\sqrt{\sin\left(\frac{\rho}{2}\right)}},
$$
reads
\begin{eqnarray}
ds^2=-4\sin\left(\frac{\rho}{2}\right)dt^2+d\rho^2+4\sin\left(\frac{\rho}{2}\right)\cos^2\left(\frac{\rho}{2}\right)d\varphi^2,
\end{eqnarray}
has a mass (\ref{solitonicmass}) given by
\begin{eqnarray}\label{massolsol4}
{\cal M}_{\tiny{\mbox{sol}}}=-\frac{4}{13}\,\beta_2\,\pi.
\end{eqnarray}

The second and third solution are identified as black string
solutions since the hyperscaling dynamical exponent $\theta=1$. The
first family of black string solution exists for $z=4$, and its line
element is
\begin{eqnarray}
ds^{2}&=&-r^{6} f(r) {dt^2}+\frac{dr^2}{r^4 f(r)}+
d\varphi^{2}, \label{HSVmetricfsoln1}\\
f(r)&=&1-\left(\frac{r_h}{r}\right)^6 , \nonumber
\end{eqnarray}
provided that the coupling constants $\beta_1$ and $\beta_2$ are
tied as
\begin{eqnarray}\label{beta1soln1}
\beta_{1}&=&-\frac{1}{3}\,\beta_2.
\end{eqnarray}
For this solution, the thermodynamics quantities are given by
\begin{eqnarray}
{\cal S}_{W}&=&-64\,\pi^{2}\,\beta_2 \,(r_h)^2,\qquad
T=\frac{3 \,(r_h)^{4}}{2\,\pi},\nonumber\\
{\cal
M}_{\tiny{\mbox{bh}}}&=&-32\,\pi\,\beta_2\,(r_h)^6,\label{entemmasssol1}
\end{eqnarray}
and, as before, it is easy to see that the first law
(\ref{firstlaw}) holds. On the other hand, using the expression
(\ref{solitonicmass}), its soliton counterpart is shown to have  a
mass
\begin{eqnarray}\label{massolsol1}
{\cal M}_{\tiny{\mbox{sol}}}=\frac{64}{9}\,\sqrt{3}\,\pi\,\beta_2.
\end{eqnarray}

The other black string solution $\theta=1$ arises for $z=1$ and is
given by
\begin{eqnarray}
ds^{2}&=&-f(r) {dt^2}+\frac{dr^2}{r^4 f(r)}+
d\varphi^{2}, \label{HSVmetricfsoln2}\\
f(r)&=&1-\left(\frac{r_h}{r}\right)^3 , \nonumber
\end{eqnarray}
with $\beta_1$ given by (\ref{beta1soln1}). For this black string
solution, the thermodynamic quantities read
\begin{eqnarray}
{\cal S}_{W}&=&16\,\pi^{2}\,\beta_2 \,(r_h)^2,\qquad
T=\frac{3 \,r_h}{4\,\pi},\nonumber\\
{\cal M}_{\tiny{\mbox{bh}}}&=&8\,\pi\,\,\beta_2\,(r_h)^3, \qquad
{\cal M}_{\tiny{\mbox{sol}}}=-{\frac {32}{27}}\,{\beta_2}\,\pi,
\label{entemmasssol2}
\end{eqnarray}
and these latter fit perfectly with the first law (\ref{firstlaw}).

The last family of solution corresponds to a Lifshitz black hole
(that is $\theta=0$) with a dynamical exponent $z=3$,
\begin{eqnarray}
ds^{2}&=&-r^{6} f(r) {dt^2}+\frac{dr^2}{r^2 f(r)}+
r^{2} d\varphi^{2}, \label{HSVmetricfsoln3}\\
f(r)&=&1-\left(\frac{r_h}{r}\right)^{4} , \nonumber
\end{eqnarray}
provided that the coupling constant $\beta_1$ is given by
(\ref{beta1soln3}). As before, the Wald entropy (\ref{waldentrop}),
the Hawking temperature (\ref{temp}) together with the masses of the
black hole and its soliton counterpart are given by
\begin{eqnarray}
{\cal S}_{W}&=&-\frac{256}{13}\,\pi^{2}\,\beta_2 \,r_h,\qquad
T=\frac{(r_h)^{3}}{\pi},\nonumber\\
{\cal
M}_{\tiny{\mbox{bh}}}&=&-\frac{64}{13}\,\pi\,\beta_2\,(r_h)^4,\quad
{\cal M}_{\tiny{\mbox{sol}}}={\frac {48}{13}}\,{2}^{2/3}
{\beta_2}\,\pi. \label{entemmasssol3}
\end{eqnarray}
In this case again, it is simple to check that the first law
(\ref{firstlaw}) is satisfied. It is somehow appealing that in new
massive gravity in three dimensions \cite{Bergshoeff:2009hq}, the
Lifshitz black hole solution also exist for a dynamical exponent
$z=3$, \cite{AyonBeato:2009nh}; this seems to confer a particular
status to the value $z=3$ concerning the Lifshitz black holes in
three dimensions for higher-order gravity theories.

We also note that there exist other solutions within the ansatz (\ref{HSVmetricf}-\ref{metricf}), but these latter have a 
vanishing mass, and hence present a little interest for our main task.

%%%%%%%%%%%%%%%%%%%%%%%%%%%%%%%%%%%%%%%%
\subsection{Generalized Cardy formula}
%%%%%%%%%%%%%%%%%%%%%%%%%%%%%%%%%%%%%%%%
In the previous sub-section, we have presented four different
families of hyperscaling violation black holes with one of them
being a Lifshitz black hole solution. In addition to the first first
law of thermodynamic (\ref{firstlaw}), the following Smarr formula
\begin{eqnarray}
{\cal M}_{\tiny{\mbox{bh}}}=\frac{1+\theta}{z+1+\theta}T\,{\cal
S}_W, \label{smarr}
\end{eqnarray}
also holds for the four classes of solutions derived previously.

Another important feature sharing by these four solutions is that
the sign of the coupling constant $\beta_2$ can always be fixed such
that the black hole mass (resp. the mass of its corresponding
soliton) is positive (resp. negative). This ensures the soliton to
be separated from the black hole spectrum by a gap.

We are now in position to propose a generalized Cardy formula for
the model considered here. We will show that the expression of the
semiclassical entropy obtained assuming that the ground state is
identified with the soliton coincides with the Wald entropy. In
order to achieve this task, we make the assumption that the dual
field theory is invariant under the following modular transformation
\begin{eqnarray}
{\cal Z}[\beta]={\cal
Z}\left[(2\pi)^{1+\frac{1+\theta}{z}}\beta^{-\frac{1+\theta}{z}}\right],
\label{modtrans}
\end{eqnarray}
where $\beta$ denotes the inverse of the temperature. This allows to
establish a relation between the partition functions at low and high
temperature regimes provided that $(1+\theta)/z>0$. Note that in the
Lifshitz case $\theta=0$, this modular transformation reduces to the
one derived in \cite{Gonzalez:2011nz} by exploiting the existence of
an isomorphism between the Lifshitz algebras with dynamical
exponents $z$ and $z^{-1}$ in $2d$. In the hyperscaling case, there
does not exist such an isomorphism. In our opinion, this is not
surprising since for the hyperscaling violation metric, in contrast
with the Lifshitz case, the effective spatial dimensionality does
not have a fixed value but explicitly depends on the theory
considered.

For each solution with the appropriate sign of the coupling constant
$\beta_2$ for which ${\cal M}_{\tiny{\mbox{bh}}}>0$ and ${\cal
M}_{\tiny{\mbox{sol}}}<0$, the existence of a gap ensures that the
partition function at low temperature is dominated by the
contribution of the ground state (the soliton),
$$
{\cal Z}[\beta]\sim \exp\left({-\beta\,{\cal
M}_{\tiny{\mbox{sol}}}}\right),
$$
and with the use of (\ref{modtrans}), we have that, at the high
temperature regime,
$$
{\cal Z}[\beta]\sim
\exp\left({-(2\pi)^{1+\frac{1+\theta}{z}}\beta^{-\frac{1+\theta}{z}}\,{\cal
M}_{\tiny{\mbox{sol}}}}\right).
$$
Hence, the asymptotic growth number of states at fixed energy ${\cal
M}_{\tiny{\mbox{bh}}}$ can be obtained from the saddle-point
approximation yielding
\begin{eqnarray}\label{generCardyquad}
{\cal S}=\frac{2\pi}{1+\theta}{\cal
M}_{\tiny{\mbox{bh}}}(1+\theta+z)\left[-\frac{{\cal
M}_{\tiny{\mbox{sol}}}(1+\theta)}{z{\cal
M}_{\tiny{\mbox{bh}}}}\right]^{\frac{z}{z+1+\theta}}.
\end{eqnarray}
Note that in the Lifshitz case ($\theta=0$), this formula reduces to
the one obtained in \cite{Gonzalez:2011nz}, and it is a matter of
check to see that for the four classes of solutions derived
previously the expression of the semiclassical entropy coincides
with the Wald entropy, ${\cal S}={\cal S}_W$. Another important
remark is that, in the case of hyperscaling violation for Einstein
gravity with a dilaton source, the Cardy formula obtained in
\cite{Shaghoulian:2015dwa} differs from (\ref{generCardyquad}) by
the change $(1+\theta)\to (1-\theta)$ which precisely corresponds to
the change of the effective spatial dimensionality
$d_{\tiny{\mbox{eff}}}$. From this last observation, one can
speculate the form of the  Cardy formula for a generic effective
spatial dimensionality $d_{\tiny{\mbox{eff}}}$ as
\begin{eqnarray}\label{generCardy}
{\cal S}=\frac{2\pi}{d_{\tiny{\mbox{eff}}}}{\cal
M}_{\tiny{\mbox{bh}}}(d_{\tiny{\mbox{eff}}}+z)\left[-\frac{{\cal
M}_{\tiny{\mbox{sol}}}\,d_{\tiny{\mbox{eff}}}}{z{\cal
M}_{\tiny{\mbox{bh}}}}\right]^{\frac{z}{d_{\tiny{\mbox{eff}}}+z}}.
\end{eqnarray}
As before, this formula can be derived assuming that the partition
function being invariant under the following modular transformation
\begin{eqnarray}
{\cal Z}[\beta]={\cal
Z}\left[(2\pi)^{1+\frac{d_{\tiny{\mbox{eff}}}}{z}}\beta^{-\frac{d_{\tiny{\mbox{eff}}}}{z}}\right].
\label{modtransgen}
\end{eqnarray}
In other words, this means that for a hyperscaling violation black
hole whose entropy scales as (\ref{scaleST}), the form of the
generalized Cardy formula will be given by (\ref{generCardy}). In
what follows, we will confirm this guess in a concrete example of
cubic gravity where the effective spatial dimensionality will be
$d_{\tiny{\mbox{eff}}}=1+3\theta$. Just to conclude this section,
let us mention that a Smarr formula can also be obtained in this
generic case from the expression of the entropy (\ref{generCardy})
together with the use of the first law yielding
\begin{eqnarray}
{\cal
M}_{\tiny{\mbox{bh}}}=\frac{d_{\tiny{\mbox{eff}}}}{z+d_{\tiny{\mbox{eff}}}}T\,{\cal
S},\label{smarrgene}
\end{eqnarray}
and generalizing the expression (\ref{smarr}) obtained in the quadratic case.

%%%%%%%%%%%%%%%%%%%%%%%%%%%%%%%%%%%%%%%%%%%%
\section{Extension to cubic gravity theory}
%%%%%%%%%%%%%%%%%%%%%%%%%%%%%%%%%%%%%%%%%%%%
In order to explore the viability of the generalized Cardy formula
(\ref{generCardy}), we now consider  the case of cubic gravity
theory in three dimensions with an action given by
\cite{Sisman:2011gz}
\begin{eqnarray}\label{actioncubic}
S[g_{\mu \nu}]&=&\int d^3x\,\sqrt{-g} \Big(\sum_{i=1}^{3} \gamma_{i}
{\cal L}_{i}\Big)\nonumber\\
&=&\int d^{3}x \sqrt{-g} {\cal L},
\end{eqnarray}
where
\begin{eqnarray*}
{\cal L}_{1}&=&R^{3},\qquad {\cal L}_{2}=RR^{ab}R_{ab},\qquad {\cal
L}_{3}=R^{ab}R_{bc}R_{\ a}^{c}.
\end{eqnarray*}
The field equations  obtained by varying the action
(\ref{actioncubic}) read \cite{Oliva:2010eb}
\begin{eqnarray}
\sum_{i=1}^{3}\gamma_{i}\,G_{\left( i\right) ab}=0,
\label{eq:cubicGrav}
\end{eqnarray}
where we have defined
\begin{eqnarray*}
G_{(1)ab} &=&3R^{2}R_{ab}+3\nabla_{p}\nabla_{q}(g_{ab}g^{pq}R^{2}
-g_{a}^{\ p}g_{b}^{\ q}R^{2})\nonumber\\
&-&\dfrac{1}{2}g_{ab}{\cal L}_{1},\\
G_{(2)ab}
&=&R_{ab}R^{cd}R_{cd}+2RR^{cd}R_{acbd}\nonumber\\
&+&\nabla_{p}\nabla _{q}(g_{ab}g^{pq}R^{cd}R_{cd}
+g^{pq}RR_{ab}\nonumber\\
&-&g_{a}^{\ p}g_{b}^{\ q}R^{cd} R_{cd} +g_{ab}RR^{pq}- g_{b}^{\
p}RR_{a}^{\ q}\nonumber\\
&-&g_{a}^{\ p}RR_{b}^{\ q})-\dfrac{1}{2}g_{ab}{\cal L}_{2},\\
G_{(3)ab} &=&3R_{acbd}R^{ec}R_{e}^{\ d}+\dfrac{3}{2}\nabla_{p}\nabla
_{q}(g^{pq}R_{a}^{\
c}R_{bc}\nonumber\\
&+&g_{ab}R^{ep}R_{e}^{\ q}-g_{b}^{\ p}R^{qc} R_{ac}-g_{a}^{\
p}R^{qc}R_{bc})\nonumber\\
&-&\dfrac{1}{2}g_{ab}{\cal L}_{3}.
\end{eqnarray*}

For an ansatz metric of the form (\ref{HSVmetricf}-\ref{metricf}),
the effective spatial dimensionality (\ref{scaleST}) will be given
by $d_{\tiny{\mbox{eff}}}=1+3\theta$. Let us check the viability of
the generalized Cardy formula (\ref{generCardy}) with the following
solution of the field equations (\ref{eq:cubicGrav}) found for
$\theta=1$ and $z=6$,
\begin{eqnarray}
ds^{2}&=&-r^{10} f(r) {dt^2}+\frac{dr^2}{r^4 f(r)}+
d\varphi^{2}, \label{HSVmetricfcubicsoln1}\\
f(r)&=&1-\left(\frac{r_h}{r}\right)^{10} , \nonumber
\end{eqnarray}
where the coupling constants are tied as
\begin{eqnarray}\label{gamma1soln1}
\gamma_{1}&=&-{\frac {11}{30}}\, {\gamma_2}-{\frac
{3}{20}}\,{\gamma_3}.
\end{eqnarray}
In this case, the thermodynamics quantities computed as before yield
\begin{eqnarray}
{\cal S}_{W}&=&2880\,{\pi}^{2} \big( 4\,{\gamma_2}+3\,{\gamma_3}
\big) (r_h)^{4} ,\qquad
T=\frac{5 \,(r_h)^{6}}{2\,\pi},\nonumber\\
{\cal M}_{\tiny{\mbox{bh}}}&=&2880\,{\pi} \big(
4\,{\gamma_2}+3\,{\gamma_3} \big)
(r_h)^{10},\label{entemmasscubicsol1}
\end{eqnarray}
and it is simple to check that the first law (\ref{firstlaw}) is
satisfied. On the other hand, for the corresponding soliton solution
obtained from (\ref{HSVmetricsoliton}) with $\alpha=10$, the
variation of the Noether potential together with the surface term
take the following forms
\begin{eqnarray}
\Delta K^{{r}{t}}&=&-{\frac {144}{5}}\, \left( 3\,{\gamma_3}
+4\,{\gamma_2} \right) \sqrt [3]{5}
,\nonumber\\
\int_{0}^{1}\!\!\!\!d{s}\,\Theta^{{r}}&=&-{\frac {288}{5}}\,
\left(3\,{\gamma_3}+4\,{\gamma_2} \right) \sqrt [3]{5},
\end{eqnarray}
giving a unique value for the mass of the soliton, independent of
any integration constant,
\begin{eqnarray}\label{massolcubicsol1}
{\cal M}_{\tiny{\mbox{sol}}}=-{\frac {864}{5}}\,\sqrt [3]{5}\,\pi\,
\left( 3\,{\gamma_3}+4\,{\gamma_2} \right).
\end{eqnarray}
As in the quadratic case, there exist a choice of the coupling
constants given by  $3\,{\gamma_3}+4\,{\gamma_2}>0$ which ensures
the mass of the black hole to be positive, and at the same time the
mass of its soliton counterpart turns to be negative. Finally, it is
a matter of check to see that the generalized Cardy formula
(\ref{generCardy}) with $d_{\tiny{\mbox{eff}}}=1+3\theta=4$ implies
as excepted that  ${\cal S}={\cal S}_W$ for the the cubic solution, and the Smarr formula (\ref{smarrgene}) is also satified.
%%%%%%%%%%%%%%%%%%%%%%
\section{Discussion}
%%%%%%%%%%%%%%%%%%%%%%

The aim of this paper was to guess the possible form for  a
generalized Cardy formula in the case of hyperscaling violation
metric in three dimensions. In order to achieve this task, we have
first stressed that, in contrast with the Lifshitz case, the
effective spatial dimensionality defined by the scaling of the
entropy in term of the temperature (\ref{scaleST})  does not take a
fixed value but clearly depends on the gravity theory considered.
For example, in the standard Einstein gravity case
$d_{\tiny{\mbox{eff}}}=1-\theta$ while for quadratic corrections we
have $d_{\tiny{\mbox{eff}}}=1+\theta$ and in the cubic case,
$d_{\tiny{\mbox{eff}}}=1+3\theta$. From these observations, we have
posited the possible form of the generalized Cardy formula in term
of the effective spatial dimensionality $d_{\tiny{\mbox{eff}}}$. As
in the Lifshitz case, the ground state is provided by the soliton
which is separated from the black hole spectrum by a gap. We have
checked the validity of this formula in different examples in the
case of quadratic and cubic gravity theories while in the standard
Einstein case our formula reduces to the one proposed in
\cite{Shaghoulian:2015dwa}. In all these examples, there exist a
choice of the coupling constants that ensures the black hole mass
(resp. the soliton mass) to be positive (resp. to be negative) which
in turn guarantees the existence of a gap in the spectrum.

Nevertheless, in contrast with the Lifshitz case where the Cardy
formula was shown to arise as a consequence of the isomorphism in
two dimensions between the Lie algebras with dynamical exponents $z$
and $z^{-1}$ \cite{Gonzalez:2011nz}, in the present case, we do not
have such an argument to justify the modular transformation
(\ref{modtransgen}). Hence, a natural extension of this work will be
to look for a justification of the duality between the low and high
temperature regimes. The exploration of new solutions can also be
interesting in order to consolidate the validity of
(\ref{generCardy}). For example, in the quadratic case, there exist
most general class of solutions within the ansatz (\ref{HSVmetricf})
without imposing the form (\ref{metricf}) to the metric function. In
fact, the first class of string solution
(\ref{HSVmetricfsoln1}-\ref{beta1soln1}) with $\theta=1$ and $z=4$
can be promoted to a two-parametric solution as
\begin{eqnarray}
ds^{2}&=&-r^{6} f(r) {dt^2}+\frac{dr^2}{r^4 f(r)}+
d\varphi^{2}, \label{HSVmetricfsolution2-par}\\
f(r)&=&1+{\frac {a}{{r}^{2}}}+{\frac {{a}^{2}}{3\,{r}^{4}}}+{\frac
{b}{{r}^{ 6}}}, \nonumber
\end{eqnarray}
where $a$ and $b$ are two integration constants. Denoting by $r_h$
the location of the horizon
$$(r_h)^{2}=\frac{1}{3}\,\sqrt [3]{\left({a}^{3}-27\,b\right)}-\frac{a}{3},$$
the expressions of the Wald entropy and temperature read
\begin{eqnarray}
{\cal S}_{W}=-{\frac {64 {\pi }^{2}}{3}}\,{\beta_2}\, \left(
3\,{r_h}^{2}+a\right) ,\quad T={\frac { \left( 3\,{r_h}^{2}+a\right)
^{2}}{6\,\pi }}, \label{entemmasssolution}
\end{eqnarray}
while the quasilocal mass which is compatible with the first law
involves as well the two integration constants as
\begin{eqnarray}
{\cal M}_{\tiny{\mbox{bh}}}={\frac {32\pi}{27}}\,{\beta_2}\, \left(
27\,b-{a}^{3} \right) =-{\frac {32\pi}{27}}\,{\beta_2}\left(
3\,{r_h}^{2}+a\right)^{3}.
\end{eqnarray}
This dependence of the mass with respect to the two integration
constants is similar with the situation occurring with the $z=1$ AdS
black hole solution of new massive gravity \cite{Oliva:2009ip} where
the solution is also two-parametric and, where the two integration
constants contribute to the expression of the mass
\cite{Perez:2011qp}. On the other hand,  the soliton counterpart of
the solution (\ref{HSVmetricfsolution2-par}) is shown to have a mass
given by (\ref{massolsol1}) and again, it is a simple exercise to
check the validity of the Cardy formula (\ref{generCardyquad}-\ref{generCardy}) in
this case.

Finally, another interesting task to be realized will be to explore
the charged version of the solutions found here in order to propose
as well a charged version of the Cardy formula.

%%%%%%%%%%%%%%%%%%%%%%%%%%%%%%%%%%%%%%%%%%%%%%%%%%%%%%%%%%%%%%%%%%%%%%%%%%%%%%%%%%%%%%%%%%%%%%%%
\begin{acknowledgments}
We thank Julio Oliva and Edgar Shaghoulian for
useful discussions.  This work is partially supported by grant
1130423 from FONDECYT and from CONICYT with grants 21120271 and
21130136. This project is also partially funded by Proyectos
CONICYT- Research Council UK - RCUK -DPI20140053.
\end{acknowledgments}
%%%%%%%%%%%%%%%%%%%%%%%%%%%%%%%%%%%%%%%%%%%%%%%%%%%%%%%%%%%%%%%%%%%%%%%%%%%%%%%%%%%%%%%%%%%%%%%%%%

%%%%%%%%%%%%%%%%%%%%%%%%%%%

%%%%%%%%%%%%%%%%%%%%%%%%%%%


\begin{thebibliography}{99}
%%%%%%%%%%%%%%%%%%%%%%%%%%%

\bibitem{Maldacena:1997re}
  J.~M.~Maldacena,
  %``The Large N limit of superconformal field theories and supergravity,''
  Adv.\ Theor.\ Math.\ Phys.\  {\bf 2}, 231 (1998).
  %%CITATION = HEP-TH/9711200;%%

\bibitem{Gouteraux:2011ce}
  B.~Gouteraux and E.~Kiritsis,
  %``Generalized Holographic Quantum Criticality at Finite Density,''
  JHEP {\bf 1112}, 036 (2011)
  [arXiv:1107.2116 [hep-th]].
  %%CITATION = ARXIV:1107.2116;%%

\bibitem{Sachdev:2012dq}
  S.~Sachdev,
  %``The Quantum phases of matter,''
  arXiv:1203.4565 [hep-th].
  %%CITATION = ARXIV:1203.4565;%%


\bibitem{Charmousis:2012dw}
  C.~Charmousis, B.~Gouteraux and E.~Kiritsis,
  %``Higher-derivative scalar-vector-tensor theories:
  black holes, Galileons, singularity cloaking and holography,''
  JHEP {\bf 1209}, 011 (2012)
  [arXiv:1206.1499 [hep-th]].
  %%CITATION = ARXIV:1206.1499;%%


\bibitem{Kachru:2008yh}
  S.~Kachru, X.~Liu and M.~Mulligan,
  %``Gravity Duals of Lifshitz-like Fixed Points,''
  Phys.\ Rev.\ D {\bf 78}, 106005 (2008).
  %%CITATION = ARXIV:0808.1725;%%


\bibitem{Taylor:2008tg}
  M.~Taylor,
  %``Non-relativistic holography,''
  arXiv:0812.0530 [hep-th].
  %%CITATION = ARXIV:0812.0530;%%

\bibitem{AyonBeato:2009nh}
  E.~Ayon-Beato, A.~Garbarz, G.~Giribet and M.~Hassaine,
  %``Lifshitz Black Hole in Three Dimensions,''
  Phys.\ Rev.\ D {\bf 80}, 104029 (2009)
  [arXiv:0909.1347 [hep-th]].
  %%CITATION = ARXIV:0909.1347





\bibitem{Pang:2009pd}
  D.~-W.~Pang,
  %``On Charged Lifshitz Black Holes,''
  JHEP {\bf 1001}, 116 (2010)
  [arXiv:0911.2777 [hep-th]].
  %%CITATION = ARXIV:0911.2777;%%

\bibitem{Maeda:2011jj}
  H.~Maeda and G.~Giribet,
  %``Lifshitz black holes in Brans-Dicke theory,''
  JHEP {\bf 1111}, 015 (2011)
  [arXiv:1105.1331 [gr-qc]].
  %%CITATION = ARXIV:1105.1331;%%
  %13 citations counted in INSPIRE as of 04 Jan 2015



\bibitem{Ayon-Beato:2015jga}
  E.~Ayón-Beato, M.~Bravo-Gaete, F.~Correa, M.~Hassaïne, M.~M.~Juárez-Aubry and J.~Oliva,
  %``First law and anisotropic Cardy formula for three-dimensional Lifshitz black holes,''
  Phys.\ Rev.\ D {\bf 91}, no. 6, 064006 (2015)
  [arXiv:1501.01244 [gr-qc]].
  %%CITATION = ARXIV:1501.01244;%%

\bibitem{Bravo-Gaete:2015xea}
  M.~Bravo-Gaete and M.~Hassaine,
  %``Thermodynamics of charged Lifshitz black holes with quadratic corrections,''
  Phys.\ Rev.\ D {\bf 91}, no. 6, 064038 (2015)
  [arXiv:1501.03348 [hep-th]].
  %%CITATION = ARXIV:1501.03348;%%




\bibitem{Correa:2014ika}
  F.~Correa, M.~Hassaine and J.~Oliva,
  %``Black holes in New Massive Gravity dressed by a (non)minimally coupled scalar field,''
  Phys.\ Rev.\ D {\bf 89}, no. 12, 124005 (2014)
  [arXiv:1403.6479 [hep-th]].
  %%CITATION = ARXIV:1403.6479;%%


\bibitem{Alvarez:2014pra}
  A.~Alvarez, E.~Ayón-Beato, H.~A.~González and M.~Hassaïne,
  %``Nonlinearly charged Lifshitz black holes for any exponent $z>1$,''
  JHEP {\bf 1406}, 041 (2014)
  [arXiv:1403.5985 [gr-qc]].
  %%CITATION = ARXIV:1403.5985;%%



\bibitem{Dong:2012se}
  X.~Dong, S.~Harrison, S.~Kachru, G.~Torroba and H.~Wang,
  %``Aspects of holography for theories with hyperscaling violation,''
  JHEP {\bf 1206}, 041 (2012)
  [arXiv:1201.1905 [hep-th]].
  %%CITATION = ARXIV:1201.1905;%%

\bibitem{Alishahiha:2012qu}
  M.~Alishahiha, E.~O Colgain and H.~Yavartanoo,
  %``Charged Black Branes with Hyperscaling Violating Factor,''
  JHEP {\bf 1211}, 137 (2012)
  [arXiv:1209.3946 [hep-th]].
  %%CITATION = ARXIV:1209.3946;%%


\bibitem{HVMBH1} M.~Cadoni and M.~Serra, JHEP {\bf 1211}, 136 (2012)
  [arXiv:1209.4484 [hep-th]];
 %%CITATION = ARXIV:1209.4484;%%

\bibitem{HVMBH2} P.~Bueno, W.~Chemissany, P.~Meessen, T.~Ortin and C.~S.~Shahbazi,
  %``Lifshitz-like Solutions with Hyperscaling Violation in Ungauged Supergravity,''
  JHEP {\bf 1301}, 189 (2013)
  [arXiv:1209.4047 [hep-th]].
  %%CITATION = ARXIV:1209.4047;%%





\bibitem{Shaghoulian:2015dwa}
  E.~Shaghoulian,
  %``A Cardy formula for holographic hyperscaling-violating theories,''
  arXiv:1504.02094 [hep-th].
  %%CITATION = ARXIV:1504.02094;%%













\bibitem{Hassaine:2015ifa}
  M.~Hassaïne,
  %``New black holes of vacuum Einstein equations with hyperscaling violation and Nil geometry horizons,''
  Phys.\ Rev.\ D {\bf 91}, no. 8, 084054 (2015)
  [arXiv:1503.01716 [hep-th]].
  %%CITATION = ARXIV:1503.01716;%%





\bibitem{Gonzalez:2011nz}
  H.~A.~Gonzalez, D.~Tempo and R.~Troncoso,
  %``Field theories with anisotropic scaling in 2D, solitons and the microscopic entropy of asymptotically Lifshitz black holes,''
  JHEP {\bf 1111}, 066 (2011)
  [arXiv:1107.3647 [hep-th]].
  %%CITATION = ARXIV:1107.3647;%%

\bibitem{Correa:2010hf}
  F.~Correa, C.~Martinez and R.~Troncoso,
  %``Scalar solitons and the microscopic entropy of hairy black holes in three dimensions,''
  JHEP {\bf 1101}, 034 (2011)
  [arXiv:1010.1259 [hep-th]];
  %%CITATION = ARXIV:1010.1259;%%


\bibitem{Kim:2013zha}
  W.~Kim, S.~Kulkarni and S.~H.~Yi,
  %``Quasilocal Conserved Charges in a Covariant Theory of Gravity,''
  Phys.\ Rev.\ Lett.\  {\bf 111}, no. 8, 081101 (2013)
  [Phys.\ Rev.\ Lett.\  {\bf 112}, no. 7, 079902 (2014)]
  [arXiv:1306.2138 [hep-th]].
  %%CITATION = ARXIV:1306.2138;%%


\bibitem{Gim:2014nba}
  Y.~Gim, W.~Kim and S.~H.~Yi,
  %``The first law of thermodynamics in Lifshitz black holes revisited,''
  JHEP {\bf 1407}, 002 (2014)
  [arXiv:1403.4704 [hep-th]].
  %%CITATION = ARXIV:1403.4704;%%

\bibitem{Bergshoeff:2009hq}
  E.~A.~Bergshoeff, O.~Hohm and P.~K.~Townsend,
  %``Massive Gravity in Three Dimensions,''
  Phys.\ Rev.\ Lett.\  {\bf 102}, 201301 (2009)
  [arXiv:0901.1766 [hep-th]].
  %%CITATION = ARXIV:0901.1766;%%

\bibitem{Sisman:2011gz}
  T.~C.~Sisman, I.~Gullu and B.~Tekin,
  %``All unitary cubic curvature gravities in D dimensions,''
  Class.\ Quant.\ Grav.\  {\bf 28}, 195004 (2011)
  [arXiv:1103.2307 [hep-th]].
  %%CITATION = ARXIV:1103.2307;%%


\bibitem{Oliva:2010eb}
  J.~Oliva and S.~Ray,
  %``A new cubic theory of gravity in five dimensions: Black hole, Birkhoff's theorem and C-function,''
  Class.\ Quant.\ Grav.\  {\bf 27}, 225002 (2010)
  [arXiv:1003.4773 [gr-qc]].
  %%CITATION = ARXIV:1003.4773;%%

\bibitem{Oliva:2009ip}
  J.~Oliva, D.~Tempo and R.~Troncoso,
  %``Three-dimensional black holes, gravitational solitons, kinks and wormholes for BHT massive gravity,''
  JHEP {\bf 0907}, 011 (2009)
  [arXiv:0905.1545 [hep-th]].
  %%CITATION = ARXIV:0905.1545;%%



\bibitem{Perez:2011qp}
  A.~Perez, D.~Tempo and R.~Troncoso,
  %``Gravitational solitons, hairy black holes and phase transitions in BHT massive gravity,''
  JHEP {\bf 1107}, 093 (2011)
  [arXiv:1106.4849 [hep-th]].
  %%CITATION = ARXIV:1106.4849;%%

%%%%%%%%%%%%%%%%%%%%%%%%%%%
\end{thebibliography}
\end{document}